\documentclass[aip,amsmath,amssymb,reprint]{revtex4-1}
\usepackage{graphicx}
\usepackage{dcolumn}
\usepackage{bm}
\usepackage{amsmath}
\usepackage{amsfonts}
\usepackage{amssymb}
\usepackage{multirow}

\usepackage{amsthm}
\usepackage{pinlabel}
\usepackage{natbib}

\usepackage{epstopdf}
\usepackage[abs]{overpic}
\usepackage[dvipsnames]{xcolor}

\newcommand{\blue}{\color{black}}

\newcommand{\ua}{\uparrow}
\newcommand{\da}{\downarrow}
\newcommand{\no}{\nonumber}
\newcommand{\ph}{^{\phantom{\dagger}}}
\usepackage{braket}

\usepackage[normalem]{ulem}

\definecolor{specialgray}{HTML}{505050}
\definecolor{col10K}{HTML}{FFA000}
\definecolor{col300K}{HTML}{924FA4}
\definecolor{colMu}{HTML}{5278BD}
\definecolor{colMuI}{HTML}{924FA4}

\definecolor{specialgray}{HTML}{505050}
\definecolor{col10K}{HTML}{FFA000}
\definecolor{col300K}{HTML}{924FA4}
\definecolor{colMu}{HTML}{5278BD}
\definecolor{colMuI}{HTML}{924FA4}
\definecolor{newred}{HTML}{D53E4F}
\definecolor{newblue}{HTML}{5278BD}
\definecolor{newcyan}{HTML}{1EA0A0}
\definecolor{newgreen}{HTML}{5CB14E}
\definecolor{newpurple}{HTML}{924FA4}
\definecolor{newyellow}{HTML}{D1C72E}
\definecolor{neworange}{HTML}{D6923C}

\usepackage{lipsum}

\begin{document}
\preprint{AIP/123-QED}

\title[Charge Density Waves beyond the Pauli paramagnetic limit in 2D systems]{Charge Density Waves beyond the Pauli paramagnetic limit in 2D systems}

\author{Alex Aperis}
\affiliation{Department of Physics and Astronomy, Uppsala University, P.\ O.\ Box 516, SE-75120 Uppsala, Sweden}
\email{alex.aperis@physics.uu.se}
\author{Georgios Varelogiannis}
\affiliation{Department of Physics, National Technical University of Athens, GR-15780 Athens, Greece}

\vskip 0.4cm
\date{\today}

\begin{abstract}
Two-dimensional materials are ideal candidates to host Charge density waves (CDWs) that exhibit paramagnetic limiting behavior, similarly to the well known case of superconductors. 
Here we study how CDWs in two-dimensional systems can survive beyond the Pauli limit when they are subjected to a strong magnetic field by developing a
 generalized mean-field theory of CDWs under Zeeman fields that includes incommensurability, imperfect nesting and temperature effects and the possibility of a competing or coexisting Spin density wave (SDW) order. Our numerical calculations yield rich phase diagrams with distinct high-field phases above the Pauli limiting field. For perfectly nested commensurate CDWs, a $q$-modulated CDW phase that is completely analogous to the superconducting {\blue Fulde-Ferrell-Larkin-Ovchinnikov (FFLO)} phase appears at high-fields. In the more common case of imperfect nesting, the commensurate CDW groundstate undergoes a series of magnetic-field-induced phase transitions first into a phase where commensurate CDW and SDW coexist and subsequently into another phase where CDW and SDW acquire a $q$-modulation that is however distinct from the pure FFLO CDW phase. The commensurate CDW+SDW phase occurs for fields comparable to but less than the Pauli limit and survives above it. Thus this phase provides a plausible mechanism for the CDW to survive at high fields without the need of forming the more fragile FFLO phase. We suggest that the recently discovered 2D materials like the transition metal dichalcogenides offer a promising platform for observing such exotic field induced CDW phenomena.
\end{abstract}

\maketitle

\section{Introduction}

Magnetic fields are detrimental to superconductivity because they tend to break the Cooper pairs either by coupling to the orbital motion or to the spin of the electrons \cite{Maki1964}. Usually, the superconducting upper critical field is limited by the orbital effect, however the latter can be minimized for thin film geometries and generally in two-dimensional systems \cite{Fulde1973}. In such a case, the superconductor is paramagnetically, or Pauli, limited and the upper critical field for conventional superconductors can be estimated by the simple BCS relation $H_P=\Delta_0/\sqrt{2}$ with $\Delta_0$ the zero temperature and magnetic field value of the superconducting gap \cite{Clogston1962,Chandrasekhar1962}. For decades it is known  that superconductors can exceed the Pauli limit via a phase transition to a modulated state with Cooper pairs that acquire a finite momentum which is driven by the external field, the so-called FFLO phase \cite{Fulde1964,Larkin1964}.

Charge Density Waves (CDWs) are quantum states of matter that are characterized by the freezing of the conduction electron charge density into a periodic modulation pattern below a critical temperature \cite{Gruener1988}. When the spin density becomes modulated instead, one speaks of a Spin Density Wave (SDW) state \cite{Gruener1994}. Charge/Spin Density Waves are frequently encountered in the phase diagrams of correlated materials where they may also compete or coexist with superconductivity and hence they have been investigated thoroughly over the past decades \cite{Schlenker1996,Monceau2012,Johnston2010,Fradkin2015}. Due to the fact that the nesting properties of the Fermi surface are enhanced with low dimensionality, these states are most typical for quasi one-dimenional systems like e.g. the Bechgaard salts \cite{Bechgaard1980}. However, density waves have also been found to occur in many two-dimensional systems like chromium films \cite{Fawcett1988}, telourides \cite{Brouet2008} and transition metal dischalcogenides (TMD) \cite{Wilson1975,Borisenko2008}. 

Up to now, the effect of strong magnetic fields on CDW/SDW states has been a subject of studies mostly in the context of 1D organic materials \cite{Brooks2008} where it has been shown that the coupling of the magnetic field to the electron's motion can give rise to field-induced Charge \cite{Zanchi1996,Lebed2003} and Spin \cite{GorkovL.P.1984,HeritierM.1984} Density Waves. However, other field-induced phases that are unrelated to orbital effects have been observed experimentally beyond 1D in diverse systems like e.g. the phases accompanied by metamagnetic transitions in URu$_2$Si$_2$ \cite{Harrison2003,Kotetes2014,Aperis2009} and the manganites 
\cite{Cao2005,Milward2005}.

Since CDWs are spin singlet condensates, they can in principle exhibit Pauli limiting behavior in complete analogy to superconductors \cite{Brooks2008}. Usually, the paramagnetic critical field of a CDW corresponds to magnetic field values of tenths of Teslas given that the critical temperature of such condensates is quite large. The quasi-1D organic salt (Per)$_2$Au(mnt)$_2$ is a rare case of a CDW material with a relatively low $T_c$ and as a result with a Pauli limit that lies in the experimentally accessible range around 37 T. In this system, a transition to a new CDW phase for $H>H_P$ was indeed observed, in which a weaker CDW gap coexists with normal state regions. This phase survives for magnetic field values that are way above the theoretical Pauli limit of the material and therefore it was identified as the first example of a FFLO phase observed in a Charge Density Wave \cite{McDonald2004}. The interpretation of this phase as a FFLO CDW relies on the dominance of the Zeeman effect \cite{McDonald2004}, however this picture is obscured by the presence of the competing orbital effect which is generally imposed by the quasi-1D nature of the system \cite{Graf2004,Zanchi1996}. 

In order to reach to the unambiguous observation of this exotic phase, it would be desirable to be able to minimize the orbital effects. Similar to the case for FFLO superconductors, optimal experimental conditions for this purpose could be achieved by applying a magnetic field in the plane of a purely two-dimensional CDW metal \cite{Fulde1973}.
In this respect, the recently synthesized single and few layer atomically thick TMDs, could offer a promising route to tackle this problem \cite{Manzeli2017}. These novel 2D materials can display enriched properties as compared to their bulk counterparts and thus they have emerged as a  testbed for the fundamental understanding of their archetypical CDWs \cite{Rossnagel2011,Xi2015a,Zhu2017,Johannes2008,Mounet2018} and the coexisting superconductivity \cite{Ugeda2015,Manzeli2017,Bekaert2020}. For example, many TMDs at their monolayer limit exhibit an intricate spin-orbit coupling that fixes the electron spins perpendicular to the plane \cite{Manzeli2017}. As a result, this mechanism gives rise to so-called Ising superconductivity which has been observed to survive under strong in-plane magnetic fields beyond the Pauli limit \cite{Xi2015,Lu2015,Sohn2018}. Interestingly, recent experiments have provided evidence of $q$-modulated superconductivity of the FFLO type for strong fields in monolayer H-NbS$_2$ \cite{Devarakonda2019} while FFLO superconductivity has also been predicted for bilayer TMDs \cite{Liu2017}. Therefore, these materials appear as suitable candidates for probing CDWs beyond the paramagnetic limit and possibly identifying the formation of the FFLO CDW phase or, as we show here, other field induced CDW phases.

Motivated by the above experimental picture, here we revisit the problem of estimating the impact of strong magnetic fields on the CDW state. It has been pointed out that CDW and SDW states generally coexist when particle-hole asymmetry and ferromagnetism are present \cite{Varelogiannis2000}. However, in the majority of previous studies these two states are considered separately, with few exceptions as e.g. in 1D systems \cite{Zanchi1996}, despite the fact that many CDW materials are strongly correlated and in fact can host SDW phases as well \cite{Monceau2012,Butz1986,Law2017}.

Here we generalize the study of cases where CDW/SDW can coexist under Zeeman fields to any system dimensionality by formulating a suitable effective mean-field theory that takes into account both these states on the same footing. In order to include the possibility of FFLO states, we extend previous works \cite{Varelogiannis2000,Aperis2008} by allowing the possibility for our considered CDW/SDW states to acquire an incommensurate modulation, i.e. our analysis fully includes incommensurate density wave ordering. As a concrete case, we numerically solve our model self-consistently for a single-band two-dimensional system. For perfectly nested 2D systems, we find that at low temperature and for magnetic fields above the Pauli limit, the CDW undergoes a first order phase phase transition into a $q$-modulated FFLO CDW which is exactly analogous to the FFLO phase of superconductors. Interestingly, when the perfect nesting condition of the underlying Fermi surface is not satisfied, we find that instead of a FFLO CDW phase the system undergoes a transition into a phase where the commensurate CDW coexists with commensurate SDW order. Remarkably, this phase becomes energetically favorable already for fields below the Pauli limit and it survives for field strengths above it. Our calculations show that this CDW+SDW phase can survive for even higher fields by undergoing a subsequent phase transition into a $q$-modulated CDW+SDW phase that bears many similarities with an FFLO state although the modulation wavevector $q$ appears to be constant with the field strength. Overall, our findings reveal a rich phase diagram for 2D CDW systems under an in-plane magnetic field and offer qualitative predictions for our observed high-field states that could be tested e.g. in two-dimensional TMDs or other 2D systems.

\section{Method}

In this section, we will first present our generalized mean-field theory of coexisting Charge/Spin Density Waves \cite{Aperis2008}, extended to include in our study the possibility of incommensurate density waves and deviations from perfect nesting. Next, we will discuss qualitatively the FFLO CDW state in the small-$q$ modulation limit within arguments based on the fermiology of the system. In the last part of the section, we provide a discussion on the method and the toy model that we use for numerical calculations.

\subsection{Theory of coexisting CDW and SDW orders under Zeeman fields\label{theory}}

Our starting point is the generalized one-band Hamiltonian that describes interacting electrons in the presence of an external Zeeman magnetic field, ${\cal H}_{tot} = {\cal H}_0 + {\cal H}_{dw}$ with
\begin{equation}\label{mf1}
{\cal H}_0 = \sum_{{\bf k},\sigma}\left(\xi\ph_{{\bf k}}-\sigma \mu_B H\right) c^{\dagger}_{{\bf k}\sigma}c\ph_{{\bf k}\sigma}\ ,
\end{equation}
where $\xi\ph_{\bf k}$ is the electron energy dispersion, $c^{(\dagger)}_{{\bf k}\sigma}$ are electron annihilation (creation) operators at momentum ${\bf k}$ and spin index $\sigma$ and $\mu_BH$ is the Zeeman spin splitting due to a magnetic field $H$ chosen parallel to the $\hat{z}$-axis ( henceforth we set $\mu_B=1$). 

Scattering processes in the particle-hole channel with exchanged momentum ${\bf Q}$ are described by the following effective four-fermion interaction Hamiltonian,
\begin{eqnarray}\label{mfDW}
{\cal H}_{dw} = -\frac{1}{2}\sum_{\bf k,k'}\sum_{s_1,s_2,s_3,s_4}c^\dagger_{{\bf k},s_1}c\ph_{{\bf k+Q},s_2}\widetilde{V}c^\dagger_{{\bf k'+Q},s3}c\ph_{{\bf k'},s_4}
\end{eqnarray}
with spin indices $s_i=\ua\da$. The interaction potential $\widetilde{V}$ can be further separated into spin singlet and spin triplet parts \cite{Sigrist1991},
\begin{align}\no
\widetilde{V}=V^w_{{\bf k,k'+Q,k+Q,k'}}\hat{\sigma}_{0_{s_1,s_2}}\hat{\sigma}_{0_{s_3,s_4}}\\\label{mfDWint}
+V^m_{{\bf k,k'+Q,k+Q,k'}}\vec{\sigma}_{s_1,s_2}\vec{\sigma}_{s_3,s_4} \ ,
\end{align}
with $\vec{\sigma}=(\hat{\sigma}_1,\hat{\sigma}_2,\hat{\sigma}_3)$ and $\hat{\sigma}_i$ the Pauli matrices.
The effective interaction potentials $V^{w(m)}$ act in the charge (spin) density wave channel and thus can mediate particle-hole ordering, respectively. These potentials can arise from the interplay between various degrees of freedom in metals, like e.g. the electron-phonon interaction (after phonons are integrated out) and the Coulomb interaction. Their microscopic origin is not important for the phenomena that we predict here and we therefore choose to keep the discussion as generic as possible by not adopting any specific microscopic mechanism \cite{Rossnagel2011}.
Within mean-field theory the interacting Hamiltonian of Eq.\,(\ref{mfDW}) can be decoupled in different CDW (W) and SDW (M) channels by introducing the generalized order parameters,
\begin{eqnarray}\no
W_{{\bf k,k+Q},s_1,s_2} &=&\sum_{\bf k'}\sum_{s_3,s_4}V^w_{{\bf k,k'+Q,k+Q,k'}}\\
&\times &\hat{\sigma}_{0_{s_1,s_2}}\hat{\sigma}_{0_{s_3,s_4}}\langle c^\dagger_{{\bf k'+Q},s_3}c\ph_{{\bf k'},s_4}\rangle \ ,\\\no
M_{{\bf k,k+Q},s_1,s_2} &=&\sum_{\bf k'}\sum_{s_3,s_4}V^m_{{\bf k,k'+Q,k+Q,k'}}\\
&\times&\vec{\sigma}_{s_1,s_3}\vec{\sigma}_{s_2,s_4}\langle c^\dagger_{{\bf k'+Q},s_3}c\ph_{{\bf k'},s_4}\rangle \ ,
\end{eqnarray}
with charge/spin modulation wavevector ${\bf Q}$. We further focus here on conventional and isotropic CDW and SDW order parameters 
 by assuming the interaction kernels in the above as momentum independent, and choose the SDW polarization parallel to that of the applied magnetic field. With these considerations, we are left with the following two order parameters,
\begin{eqnarray}\label{meanfields1}
W&=&\sum_{{\bf k'},\sigma}V^w\langle c^\dagger_{{\bf k'},\sigma}c\ph_{{\bf k'+Q},\sigma}\rangle \ ,\\\label{meanfields2}
M&=&\sum_{{\bf k'},\sigma}V^m\sigma\langle c^\dagger_{{\bf k}',\sigma}c\ph_{{\bf k'+Q},\sigma}\rangle
\end{eqnarray}
and the resulting mean-field Hamiltonian reads \cite{Varelogiannis2000},
\begin{eqnarray}\no
{\cal H} &=& \sum_{{\bf k},\sigma}\left(\xi\ph_{{\bf k}}-\sigma H\right) c^{\dagger}_{{\bf k}\sigma}c\ph_{{\bf k}\sigma}
-\frac{1}{2}W\sum_{{\bf k},\sigma}\left(c^{\dagger}_{{\bf k}\sigma}c\ph_{{\bf k}+\bf{Q}\sigma} +
\text{H.c.}\right)\\\label{inham}
&-&\frac{1}{2}M\sum_{{\bf k},\sigma}\sigma \left(
c^{\dagger}_{{\bf k}\sigma}c\ph_{\bf{k}+\bf{Q}\sigma} + \text{H.c.}\right)+\frac{W^2}{V^w}+\frac{M^2}{V^m} \ .
\end{eqnarray}
The second and third term in the above are the mean-field Hamiltonians of CDW and SDW, respectively in analogy to the BCS theory and the last two terms that arise from the decoupling process can be understood as the energy barrier that the system has to overcome in order for condensation to be energetically favorable (see e.g. Eq.\,(\ref{fr}) below).

With the above considerations, Eq.\,({\ref{inham}}) can be compactly rewritten with the use of the following spinor,
\begin{eqnarray}\label{spinor}
\zeta^\dagger_{{\bf k},\sigma}=\frac{1}{\sqrt{2}}(c^\dagger_{{\bf k}\sigma},c^\dagger_{{\bf k+Q}\sigma}) ,
\end{eqnarray}
and the $2\times 2$ basis of $\hat{\rho}_i$ Pauli matrices as,  
\begin{equation}\label{ham2}
{\cal H} = \sum_{\bf{k},\sigma}\zeta^{\dagger}_{{\bf k},\sigma} \bigl(\gamma\ph_{\bf k}\hat{\rho}_3+\delta\ph_{\bf k}-W\hat{\rho}_1-\sigma M\hat{\rho}_1-\sigma H\bigl) \zeta\ph_{{\bf k},\sigma}\ ,
\end{equation}
where the last two terms in Eq.\,(\ref{inham}) are omitted for now.
In the above, we have decomposed the electron energy dispersion of Eq.\,(\ref{inham}) into two terms, $\xi\ph_{\bf k}=\gamma\ph_{\bf k}+\delta\ph_{\bf k}$, with the functions $\gamma\ph_{\bf k}$ and $\delta\ph_{\bf k}$ given by the relations,
\begin{eqnarray}\label{phrels}
\gamma\ph_{\bf k} = \frac{\xi_{{\bf k}}-\xi_{{\bf k+Q}}}{2}\ \ , \ \ 
\delta\ph_{\bf k} = \frac{\xi_{{\bf k}}+\xi_{{\bf k+Q}}}{2}\ .
\end{eqnarray}
Recalling the nesting condition $\xi\ph_{\bf k}=-\xi\ph_{\bf k+Q}$, $\gamma\ph_{\bf k}$ can be understood as the nested part of the bandstructure whereas $\delta\ph_{\bf k}$ as a term measuring deviations from perfect nesting. For the case of a commensurate wavevector, i.e. ${\bf Q}={\bf Q}_0$ with $2{\bf Q}_0$ a reciprocal wavevector, $\gamma\ph_{\bf k}$
is antisymmetric and $\delta\ph_{\bf k}$ is symmetric with respect to ${\bf Q}_0$-translations. In this special case, $\gamma\ph_{\bf k}$ is a particle-hole symmetric term and $\delta\ph_{\bf k}$ measures particle-hole asymmetry in the system \cite{Varelogiannis2000,Aperis2008JPCM,Aperis2010}.

The Hamiltonian in Eq.\,(\ref{ham2}) is quadratic and can be diagonalized by means of a fermionic Bogoliubov transformation that yields the four quasiparticle energy dispersions,
\begin{eqnarray}\label{polesCDW}
E_{\sigma\pm}({\bf k})=\delta_{\bf k}-\sigma H\pm\sqrt{\gamma_{\bf k}^2+(W+\sigma M)^2} \ .
\end{eqnarray}
The free energy of the system can be found from $F=-T\ln{\cal Z}$ where ${\cal Z}$ is the fermionic partition function. We find, 
\begin{eqnarray}\label{fr}
{\cal F}=\frac{W^2}{V^w}+\frac{M^2}{V^m}-\frac{T}{2}\sum_{{\bf k},\sigma}\sum_\pm\ln{\left(1+e^{-\frac{E_{\sigma\pm}({\bf k})}{T}}\right)} .\ \ \ \ 
\end{eqnarray}
By minimizing the above free energy with respect to our order parameters, i.e. taking $\vartheta{\cal F}/\vartheta W=0$ and $\vartheta{\cal F}/\vartheta M=0$, we arrive at the following set of coupled self-consistent equations,
{\blue \begin{eqnarray}\no
W=V^w&\sum_{{\bf k},\sigma}&\frac{W+\sigma M}{2\left[E_{\sigma+}({\bf k})-E_{\sigma-}({\bf k})\right]}\\\label{eqW}
&\times&\left(n_F\left[E_{\sigma-}({\bf k})\right]-n_F\left[E_{\sigma+}({\bf k})\right]\right) \ ,\\\no
M=V^m&\sum_{{\bf k},\sigma}&\frac{M+\sigma W}{2\left[E_{\sigma+}({\bf k})-E_{\sigma-}({\bf k})\right]}\\\label{eqM}
&\times&\left(n_F\left[E_{\sigma-}({\bf k})\right]-n_F\left[E_{\sigma+}({\bf k})\right]\right)\ .
\end{eqnarray}
The above equations have the interesting feature that on their right-hand-side there exist terms that are not proportional to the order parameter of the left-hand-side. They thus differ from the typical BCS equations that one would have obtained if the CDW/SDW orders were not studied on the same footing, i.e. taken separately. Setting $M=0$ on the right-hand-side of Eq.\,(\ref{eqM}), one can observe that for $W\neq 0$, the SDW order parameter on the left-hand-side can be nonzero if additionally $\delta\ph_{\bf k}\neq 0$ and $H\neq 0$. The same holds for the CDW case if we set $W=0$ on the right-hand-side of Eq.\,(\ref{eqW}) and assume $M\neq 0$, instead. Therefore, we see that CDW or SDW ordering can be induced in a system where one of them exists in the presence of finite $\delta\ph_{\bf}$ and $H$. In this sense, these four terms form a pattern of coexisting condensates \cite{Varelogiannis2000,Aperis2008JPCM,Aperis2010} and this property will be pivotal in understanding our numerical results presented below. As a crosscheck, one can show that for $\delta\ph_{\bf k}=0$, Eqs.\,(\ref{eqW}-\ref{eqM}) coincide with those obtained previously from a Green's function approach \cite{Aperis2008}.

Eqs.\,(\ref{eqW}-\ref{eqM}) can be solved iteratively to determine the corresponding values of $W,M$. This was done previously in the case of ${\bf Q}={\bf Q}_0$ and $\delta\ph_{\bf k}=0$ \cite{Aperis2008}. Here, the theory of coexisting CDWs/SDWs is extended to include incommensurability effects by allowing ${\bf Q}$ to be determined by minimizing either the free energy of Eq.\,(\ref{fr}) or the free energy difference between the condensed and the normal state, 
\begin{eqnarray}\label{dfr}
\delta{\cal F}=\frac{W^2}{V^w}+\frac{M^2}{V^m}-\frac{T}{2}\sum_{{\bf k},\sigma}\sum_\pm\ln{\frac{1+e^{-E_{\sigma\pm}({\bf k})/T}}{1+e^{- \epsilon\ph_{\sigma\pm}({\bf k})/T}}}
\end{eqnarray}
with $\epsilon\ph_{\sigma\pm}({\bf k})$ the normal state energy dispersions corresponding to setting $W=M=0$ in Eq.\,(\ref{polesCDW}). In practice, we will use Eq.\,(\ref{dfr}) since this allows to avoid cases where the condensed state is a local free energy minimum and the global minimum is achieved in the normal state.

For completeness, the induced magnetization of the system can be found by the relation ${\cal M}=-\vartheta{\cal F}/\vartheta H$ which yields,
\begin{eqnarray}\label{mag}
{\cal M}&=&\frac{\mu_B}{2}\sum_{{\bf k},\sigma}\sigma \left[n_F(E_{\sigma,-}({\bf k}))+n_F(E_{\sigma,+}({\bf k}))\right]\ .
\end{eqnarray}
The corresponding induced ferromagnetic splitting (FM), which is measured in units of energy, is found from $\widetilde{H}={\cal M}H$ as,
\begin{eqnarray}\label{F}
\widetilde{H}&=&\frac{H}{2}\sum_{{\bf k},\sigma}\sigma \left[n_F(E_{\sigma,-}({\bf k}))+n_F(E_{\sigma,+}({\bf k}))\right]\ .
\end{eqnarray}
Eqs.\,(\ref{mag}-\ref{F}) admit as input the self-consistently obtained $W,M$ and ${\bf Q}$ values.
For $W=M=0$ and sufficiently large magnetic fields so that the medium is fully polarized, Eq.\,(\ref{F}) yields $\widetilde{H}=h$ whereas it gives $\widetilde{H}<H$ in all other situations, as it should. 

It is worth pointing out that in the special case where the ordering wavevector ${\bf Q}={\bf Q}_0$ is commensurate, it is possible to work in the folded Brillouin Zone (BZ). However, since here no prior assumption regarding the commensurability of the CDW/SDW orders is made, all ${\bf k}$-sums are taken in the full (unfolded) BZ, instead.

\subsection{Qualitative discussion of the FFLO CDW state\label{cdwfflo}}

Before proceeding with the numerical solutions to our model, we will first provide a heuristic discussion on the mechanism of FFLO CDW formation by examining how the density wave wavevector ${\bf Q}$ can be affected due to changes in the topology of the underlying Fermi surface. For a weakly coupled density wave system where the momentum dependence of the effective interactions is not essential to the resulting electron-hole pairing, Eq.\,(\ref{meanfields1}) implies that the CDW order is maximized when ${\bf Q}$ is such that it satisfies the general nesting condition:
\begin{eqnarray}\label{nesting}
\xi\ph_{{\bf k},\sigma}=-\xi\ph_{{\bf k+Q},\sigma}\ ,
\end{eqnarray}
for as many ${\bf k}$-points in the Brillouin Zone as possible. In the above $\xi\ph_{{\bf k},\sigma}=\xi\ph_{{\bf k}}-\sigma H$ as in Eq.\,(\ref{mf1}). 
Without loss of generality, we can make further progress by writing ${\bf Q}$ as
\begin{eqnarray}\label{Q}
{\bf Q}={\bf Q}_0+{\bf q}
\end{eqnarray}
where ${\bf Q}_0$ is a commensurate wavevector as discussed previously and ${\bf q}$ measures possible deviations from incommensurability. Next, we substitute Eq.\,(\ref{Q}) into Eq.\,(\ref{nesting}) and Taylor expand both sides of Eq.\,(\ref{nesting}) around ${\bf q}=0$. Keeping only ${\cal O}(q)$ terms we have,
\begin{eqnarray}\label{CDWnestCond1}\no
\xi_{{\bf k}}+\xi_{{\bf k+Q}_0}-2\sigma\,H= -{\bf q}\cdot(\nabla_{\bf k}\xi_{{\bf k+Q}_0})\ .
\end{eqnarray}
From Eq.\,(\ref{phrels}) and the related discussion and assuming for simplicity that the ${\bf Q}_0$-symmetric term is independent of ${\bf k}$, i.e. $\delta\ph_{\bf k}=-\mu$ with $\mu$ the chemical potential, we arrive at the relation,
\begin{eqnarray}\label{CDWnestCond2}
q=\frac{2(\mu+\sigma\,H)}{\upsilon_{F} x}=(\mu+\sigma\,H)\alpha\ ,
\end{eqnarray}
where $\alpha=\frac{2}{\upsilon_{F} x}$, $\upsilon_F$ is the Fermi velocity and $x=\cos{\theta}$ with $\theta$ the angle between ${\bf q}$ and the Fermi wavevector ${\bf k}_F$. The above result provides a qualitative estimate of $q$ in the limiting case of a constant DOS at the Fermi level or in the case of a one-dimensional system \cite{Zanchi1996}. For two-dimensional systems like the ones we are interested in here, $\xi\ph_{\bf k}$ can generally be quite anisotropic in momentum space. As a result, $\nabla_{\bf k}\xi_{\bf k}\neq\upsilon_F$ and the optimal choice of $\alpha$ depends on the BZ direction with the highest DOS near the Fermi level and should be obtained numerically by employing the theory presented in the previous section. The simplifications used in this section are useful to reach to a qualitative understanding, more realistic situations are discussed in Sec.\,\ref{results} where numerical results are presented.

Eq.\,(\ref{CDWnestCond2}) is the CDW analogue to the  celebrated FFLO result for the case of superconductors \cite{Fulde1964,Larkin1964}. Similar to the superconducting case, it states that it is possible for the particle-hole pairs of a CDW state to acquire an extra $q$-modulation which is linearly proportional to the external magnetic field. In other words, it is possible for a CDW to become incommensurate in order to survive at high enough magnetic fields. The basic difference with the superconducting case is that $|q|$ here is a function of spin, thus allowing for a possible phase between the charge density of each spin species, and concomitantly the induction of a spin density wave \cite{Zanchi1996}. One can see this effect clearer if we write down the equations for the modulation of charge and spin in real space,
\begin{eqnarray}\no
\rho_c({\bf r})&\propto&\sum_\sigma \cos{\left[({\bf Q}_0+(\mu+\sigma\,H)\alpha)\,{\bf r}\right]}\\\label{rhoc}
&=&2\cos{\left[({\bf Q}_0+\alpha\,\mu)\,{\bf r}\right]}\cos{\left[(\alpha\,H)\,{\bf r}\right]}\ ,\\\no
\rho_s({\bf r})&\propto&\sum_\sigma \sigma \cos{[({\bf Q}_0+(\mu+\sigma\,H)\alpha)\,{\bf r}]}\\\label{rhos}
&=&2\cos{\left[({\bf Q}_0+\alpha\,\mu)\,{\bf r}+\frac{\pi}{2}\right]}\cos{\left[(\alpha\,H)\,{\bf r}-\frac{\pi}{2}\right]}.\  \ \ \ 
\end{eqnarray}
From the above equations one immediately observes that for $H\neq 0$ the modulated part of the spin density, $\rho_s$, is nonzero signaling the induction of a SDW. As seen from Eqs.\,(\ref{CDWnestCond2}) and (\ref{rhoc}-\ref{rhos}) terms that destroy the perfect nesting condition of Eq.\,(\ref{nesting}),  like e.g. a chemical potential, may also lead to incommensurate CDWs, as is generally expected. The relations in Eqs.\,(\ref{rhoc}-\ref{rhos}) constitute the generalization of the so-called double cosine phase that has been discussed in 1D systems \cite{Grigoriev2005}.

We note here that the situation discussed in this section concerns a system where only CDW ordering is assumed in contrast to the more complete theory that we developed in the previous section where both CDW and SDW orders are included on equal footing. In this respect, the mechanism of SDW induction due to the incommensurate FFLO CDW that is implied by Eq.\,(\ref{rhos}) is different from the field-induced coexistence of CDW+SDW states that we discussed in relation to Eqs.\,(\ref{eqW}-\ref{eqM}). For example, this difference can be observed from the fact that Eq.\,(\ref{rhos}) gives an induced SDW even when $\mu=0$ whereas Eqs.\,(\ref{eqW}-\ref{eqM}) indicate that $\mu\neq 0$ is necessary for this to happen. As we will show below our direct numerical solutions verify the latter physical picture.

\subsection{Details for the exact numerical solution to the model\label{details}}

In this section we describe the procedure for the numerical solution to the model introduced in Sec.\,\ref{theory}. Given a specific electron energy dispersion, a magnetic field strength and interaction potentials, Eq.\,(\ref{inham}) contains three unknowns: the density wave gaps $W,M$ and the ordering wavevector ${\bf Q}$. Our method for obtaining an exact solution to Eq.~(\ref{inham}) consists of simultaneously minimizing the free energy difference given by Eq.~(\ref{dfr}) with respect to the gaps $W,M$ \textit{and} the optimal wavevector ${\bf Q}$. Instead of working with ${\bf Q}$, we decompose it as in Eq.\,(\ref{Q}) and, noting that ${\bf Q}_0$ is fixed by the choice of the underlying bandstructure, we are left with ${\bf q}$ as the unknown wavevector, instead. For our calculations, we assume an electron energy dispersion given by a square lattice tight-binding (TB) model with nearest neighbor hopping energy, $t$, and chemical potential $\mu$, 
\begin{equation}
\xi\ph_{\bf k}=-t\left(\cos{k_x}+\cos{k_y}\right)-\mu\ .
\end{equation}
For this dispersion, ${\bf Q}_0=(\pi,\pi)/a$, with $a$ the lattice constant (here $a=1$). {\blue Inclusion of longer-range hopping, i.e. to next-nearest neighbors etc, is allowed by our theory. Such terms would contribute to $\delta_{\bf k}$ since they generally lead to imperfect nesting. Therefore, they would generally promote the coexistence of CDW+SDW under applied magnetic fields if of course they are not so strong so as to destroy the CDW groundstate altogether. Here $\delta\ph_{\bf k}=-\mu$ is chosen for simplicity as discussed below.}

In all our calculations, we set $t=1$ and vary temperature, $T$, magnetic field strength, $H$, and $\mu$ for a given choice of $V^{w(m)}$. All quantities are measured in units of $t$. Numerical solutions are achieved by employing a parallelized numerical code that iteratively solves the set of coupled self-consistent equations (\ref{eqW}-\ref{eqM}) on a $64\times 64$ $k$-grid in the full BZ for different values of the wavevector ${\bf q}$ that are taken from a $32\times 32$ $q$-grid in the irreducible wedge of the BZ. In this way, for each set of parameters ($T,H,\mu$) we calculate $W,M$ and the corresponding free energy at every ${\bf q}$. The physical solution that is kept is the one that minimizes $\delta{\cal F}$. 
Given the numerical complexity of the involved calculations, we chose to restrict our tight-binding model only to nearest neighbors so that the optimal ${\bf q}$ (when it is found to be nonzero) always forms a 45$^0$ angle with ${\bf Q}_0$, i.e. it always points at the Van Hove points where the DOS is maximal. Technically, this allows us to focus only on determining the amplitude, $q$, of the wavevector ${\bf q}$. 
The final density wave solution that is obtained is a superposition of harmonics with ${\bf Q}=(\pm Q_0\pm q,\pm Q_0)$ and ${\bf Q}=(\pm Q_0,\pm Q_0\pm q)$. 

\section{Results-Discussion\label{results}}

We have repeated the computational procedure of Sec.\,\ref{details} for several sets of $(V^w,V^m)$ values and found that depending on the relative strength of these potentials the ground state solution can be either a CDW or a SDW state, as expected. Focusing on cases with a commensurate CDW as groundstate solution, there exists a wide range of values for the ratio $V^w/V^m$ where the phase transition phenomena that we report in this section can be triggered by varying ($T,H,\mu$). As a general trend, the system becomes more susceptible to such phenomena as the ratio $V^w/V^m$ approaches unity due to the interplay between CDW and SDW orders becoming more pronounced. As a representative example, here we report results for ($V^w=1.5,V^m=1.2$). This choice corresponds to a system with a CDW groundstate but enhanced effective interactions in the SDW channel. For example, such a situation could arise in a material where the electron-phonon interaction is sufficient to drive the system to a CDW instability but Coulomb interactions are nevertheless pronounced, as is typical for two-dimensional systems.

Fig.\,\ref{fig1}(a) shows the calculated $H$-$T$ phase diagram of a CDW insulator for $\mu=0$. For sufficiently low fields and high temperature or high enough temperature and low fields, the CDW order gives way to the normal state through a second order phase transition (marked by solid lines). At low temperatures the transition from this commensurate CDW to the normal state becomes first order (marked by dashed lines). Had we not included incommensurability effects in our theory, this would have been a typical phase diagram of a Pauli limited spin singlet condensate similar to e.g. that of an s-wave superconductor \cite{}. However, in our case, we find that at very low temperatures a first order transition from commensurate to incommensurate CDW takes place. The latter phase is continuously suppressed for higher fields and disappears through a second order phase transition in accordance with what is expected for a FFLO CDW phase \cite{Fulde1964,Zanchi1996}.
\begin{figure}[t!]\begin{center}
\includegraphics[width=1.0\linewidth]{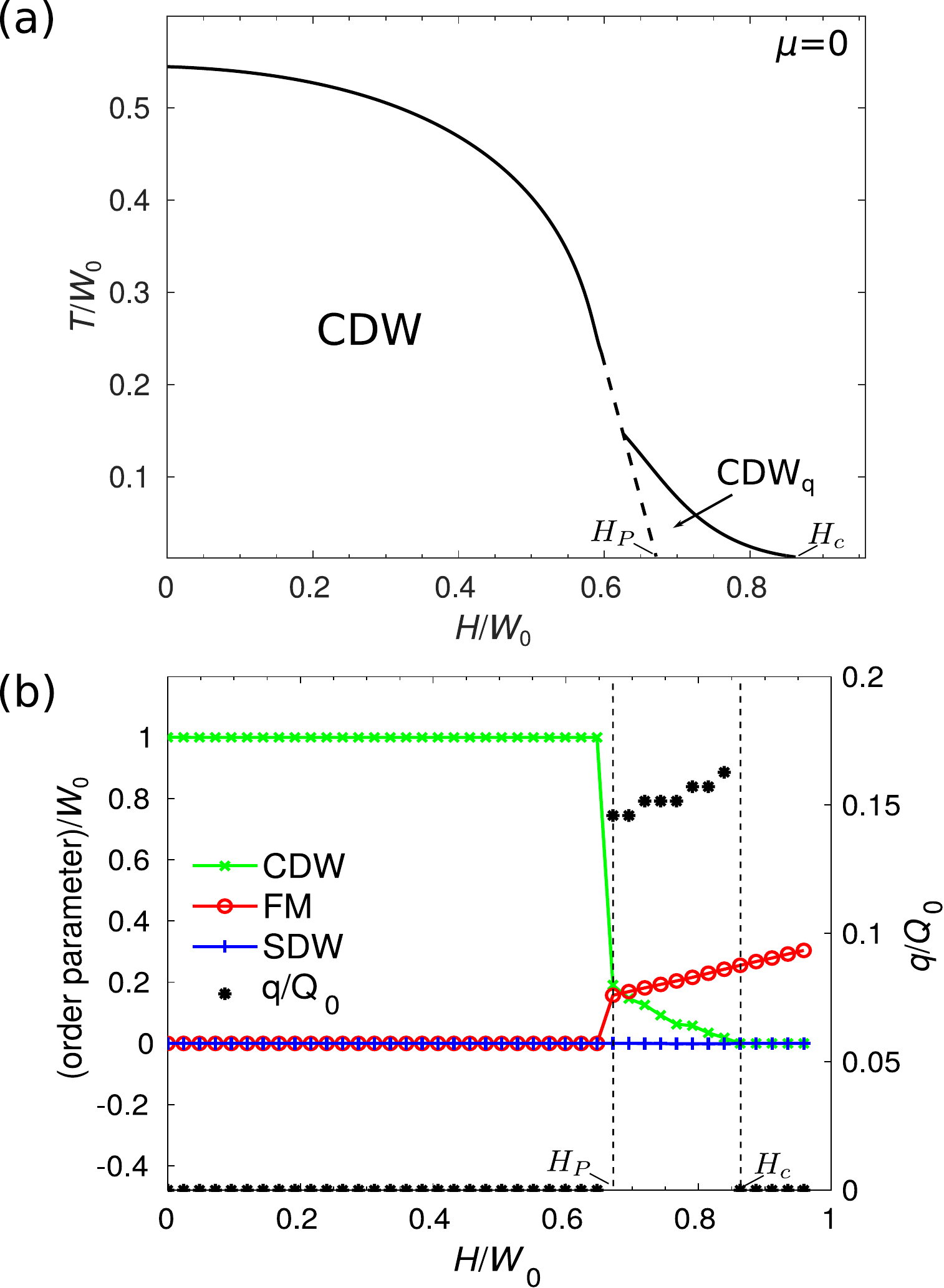}
\caption{Results for $\mu=0$. (a) Calculated H-T phase diagram. Second and first order phase transitions are marked with solid and dashed lines, respectively. {\blue Both axes are normalized by $W_0$ which is the value of the CDW gap for $T=H=0$.} The FFLO CDW phase is indicated with CDW$_q$ since there $q$ is finite as shown below. (b) Left axis: calculated magnetic field dependence of the order parameters $W$ (green), $\widetilde{H}$ (red) and $M$ (blue), {\blue normalized by $W_0$}, at $T=0$. Right axis: calculated magnetic field dependence of the amplitude of the incommensurability wavevector $q$ normalized by the amplitude of the commensurate wavevector, $Q_0$, shown with black sympols. {\blue In both (a) and (b) the Pauli and the upper critical field are indicated by $H_P$ and $H_c$, respectively.} \label{fig1}}
\end{center}\end{figure}

To gain more insight on this phase, we show in Fig.\,\ref{fig1}(b) the calculated zero-temperature, magnetic-field dependence of all order parameters in our theory. In the same Figure, is shown the amplitude of the calculated incommensuration wavevector normalized by the respective value of the commensurate wavevector, $q/Q_0$ (black star symbols). The left axis in this Figure measures the order parameter value (in units of $t$) and the right axis the $q/Q_0$ ratio. The values of the x-axis are normalized as $H/W_0$, where $W_0$ is the CDW gap value for $T=0,H=0$. We find that there exists a range of magnetic field values where the CDW (green line) becomes incommensurate with $Q=Q_0+q$ and $q\propto H$. The transition to this phase is first order and occurs at $H_P\approx0.65W_0$ which is quite close to the expected Pauli limit for systems with isotropic Fermi surfaces, $H^{iso}_P\approx0.707W_0$ \cite{Clogston1962,Chandrasekhar1962}, thus the deviation of our result from the latter value is attributed to the anisotropic underlying bandstructure. The high field CDW phase survives for field strengths above the Pauli limit $H_c=0.86W_0>H_P$ before it is destroyed via the second order phase transition. For $H<H_P$, the CDW is commensurate and has a full gap over the Fermi surface. This can be deduced by the fact that the induced ferromagnetic splitting (red line) is zero in this region. In the CDW$_q$ phase, the modulated CDW order allows for a coexistence of gapped CDW and normal state regions. The latter are polarized by the field leading to a finite ferromagnetic splitting.  The above features are in perfect agreement with the predictions from the FFLO theory for $s$-wave superconductors \cite{Fulde1964,Fulde1973}, thus in this low temperature-high field region the solution is a typical FFLO CDW phase as discussed qualitatively in Sec.\,\ref{cdwfflo}. However, in contrast to Eq.\,(\ref{rhos}) which predicts the emergence of an accompanying SDW in this phase, for the case of $\mu=0$ we find that the SDW order is absent (blue line). This is despite the fact that interactions in the SDW channel are included in the theory.

The phase diagram of Fig.\,\ref{fig1}(a) is significantly changed when $\mu\neq 0$. Here we focus on the case where $\mu=0.1$. This value is small enough so as not to influence the ground state CDW solution, i.e. $W_0(\mu=0.1)=W_0(\mu=0)$ and concomitantly the expected value of $H_P$ is the same as for $\mu=0$. Results for $\mu=0.1$ are presented in Figs.\,\ref{fig2}(a),(b) where the notations follow those of Figs.\,\ref{fig1}(a),(b). As shown in Fig.~\ref{fig2}(a), new transitions appear which notably involve a finite SDW order. As discussed in Sec.\,\ref{theory}, the presence of the SDW order can be understood by the fact that the order parameters $W,M$ always coexist when $\mu\neq0$ and $H\neq0$ thus forming a pattern of coexisting CDW+SDW condensates \cite{Varelogiannis2000,Aperis2008JPCM,Aperis2010}. A consequence of this mechanism is that at sufficiently high temperatures where thermal quasiparticle excitations above the CDW gap become possible, the presence of the applied magnetic field results in the induction of a weak SDW order which is labelled as CDW+wSDW phase in Fig.\,\ref{fig1}(a). This phase, appears as a smooth crossover which is indicated by the dashed-dotted line in Fig.\,\ref{fig1}(a).

Much more interesting is the low temperature-high field part of the phase diagram that exhibits cascades of new magnetic field induced transitions. In this case, as the field increases the system undergoes a first order phase transition from commensurate CDW order to a phase where commensurate CDW and SDW coexist. In the CDW+SDW phase $W$ and $M$ have similar magnitudes as can be seen from Fig.\,\ref{fig2}(b). For zero $T$, this transition takes place at $H_{c_1}\approx0.53W_0$, a value that is almost 20\% lower than the expected Pauli limiting field. Notably, the reduced $H_{c_1}$ makes this phase more easily accessible by experiment as compared to the FFLO CDW case. This coexisting CDW+SDW phase is also a superposition of gapped and normal state regions, similar to the FFLO CDW phase as can be inferred by the built up of finite ferromagnetic splitting (see Fig.\,\ref{fig2}(b)). {\blue These gapless Fermi surface portions could manifest in experiments as resistance drops when the external magnetic field sweeps across the CDW+SDW phase transition, similar to the transport anomalies observed e.g. in TaS$_2$ \cite{Ribak2017}. However, since a large fraction of the carriers is frozen into the CDW and SDW condensates, the anticipated resistance drop will be less than what one would observe in the normal state of the metal. The }
CDW+SDW phase survives for fields up to $H_{c_2}\approx 0.79W_0$, thus already surpassing the Pauli limit by 18\%. These characteristics resemble those of the FFLO CDW phase. However, our field induced CDW+SDW phase is markedly different; it exhibits no $q$-modulation, therefore it is \textit{commensurate}, and the transition out of this phase which takes place at H$_{c_2}$ is also first order. In fact, from Fig.\,\ref{fig2}(b) one can see that this phase appears to be bounded by two metamagnetic transitions at H$_{c_1}$ and H$_{c_2}$.
\begin{figure}[t!]\begin{center}
\includegraphics[width=1\linewidth]{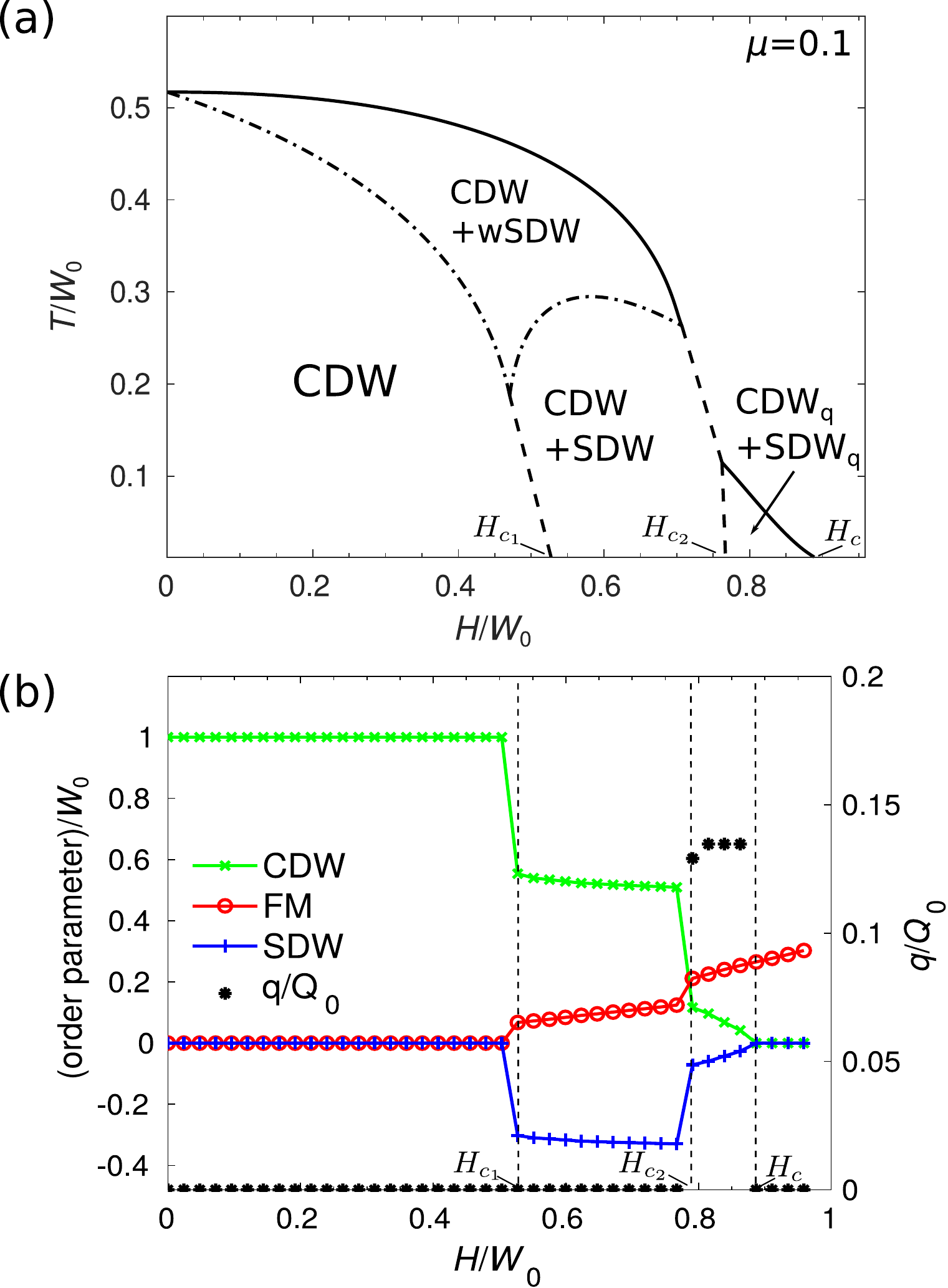}
\caption{Results for $\mu=0.1$. (a) Calculated H-T phase diagram. Second and first order phase transitions are marked with solid and dashed lines, respectively. {\blue Both axes are normalized by $W_0$ which is the value of the CDW gap for $T=H=0$.} The dotted-dashed line marks a crossover region where a very weak SDW order parameter (wSDW) is finite due to thermal excitations above the CDW gap. CDW+SDW indicates the coexistence phase of commensurate charge and spin density waves, CDW$_q$+SDW$_q$ indicates the coexistence of FFLO CDW and SDW phases. (b) Left axis: calculated magnetic field dependence of the order parameters $W$ (green), $\widetilde{H}$ (red) and $M$ (blue), {\blue normalized by $W_0$}, at $T=0$. Right axis: calculated magnetic field dependence of the amplitude of the incommensurability wavevector $q$ normalized by the amplitude of the commensurate wavevector, $Q_0$, shown with black sympols. {\blue In both (a) and (b) the critical fields of the transition into the commensurate CDW+SDW phase, the FFLO CDW+SDW phase and the upper critical field are indicated by $H_{c_1}$, $H_{c_2}$ and $H_c$, respectively.}\label{fig2}}
\end{center}\end{figure}

Remarkably, Fig.\,\ref{fig2}(a) shows that depending on the temperature, the transition at $H_{c_2}$ may either be to the normal state or to yet another ordered phase. At low $T$ and for $H>H_{c_2}$, the system enters into another coexistence phase where both CDW and SDW are modulated with an additional wavevector $q$. By acquiring this incommensurate modulation, the CDW state survives for up to even higher magnetic fields and eventually disappears at $H_c\approx 0.89W_0$ through a second order  transition (e.g. see Fig.\,\ref{fig2}(a),(b)). This FFLO phase differs from the one found for $\mu=0$ where there is no SDW order. Moreover, as can be seen in Fig\,\ref{fig2}(b), there is no clear linear correlation between $q$ and $H$. In the commensurate coexistence phase, the energy gaps $W,M$ are almost half of the $W_0$ value. In contrast, in the modulated coexistence phase, the gaps are an order of magnitude smaller than $W_0$. Therefore this latter phase could be particularly diffficult to observe experimentally. This phase is also expected to be fragile against the presense of impurities and therefore very clean samples would be required for its experimental detection. It is also worth noting that for $V^m\rightarrow V^w$, the pure FFLO phase becomes less energetically favorable as compared to the commensurate coexistence phase, even when $\mu\rightarrow 0$ \cite{Aperis2008}. 

The results of Fig.\,\ref{fig2} show that in CDW systems with imperfectly nested Fermi surfaces, the commensurate CDW+SDW phase is energetically stable over a much wider range of temperatures and magnetic field strengths as compared to the $q$-modulated CDW+SDW phase. In addition, by entering into the former phase a CDW can exceed the Pauli limit without forming the more fragile FFLO phase. This means that in CDW insulators that survive beyond the Pauli limit, the high-field phase can be a mixture of CDW+SDW orders without any additional modulation of the density wave wavevector as compared to that of the ground state.

Our findings also indicate that the mechanism for the coexistence of CDW and SDW inside the high-field phase of CDW systems is not simply due to  incommensurability effects that are driven by $q\neq0$ as Eq.\,(\ref{rhos}) implies and as was suggested in the case of 1D systems \cite{Zanchi1996}. Instead, the driving mechanism for the induction of such a phase is more general and is due to deviations from perfect nesting in the underlying bandstructure that enforce the coexistence of CDW and SDW orders in the presence of a magnetic field, as discussed in Sec.\,\ref{theory} and previously\cite{Varelogiannis2000}. These deviations, here expressed as $\mu\neq0$, are always present when the term $\delta\ph_{\bf k}$ of Eq.\,(\ref{phrels}) is nonzero, a situation that is actually the most common in real 2D systems. 

{\blue It is easy to observe that the only momentum dependent quantities entering in Eqs.\,(\ref{eqW}-\ref{eqM}) are  the poles of Eq.\,(\ref{polesCDW}). The form of these  poles does not depend explicitly on the choice of a specific underlying TB model but it is dictated by the properties of our assumed density wave order parameters. In this respect, and similarly to the single-{\bf Q} case that we assume here, multi-{\bf Q} CDW and SDW orders can coexist under the application of Zeeman fields when the corresponding $\delta\ph_{\bf k}$ is nonzero, as well. Additionally, the same mechanism can lead to the coexistence of CDWs and SDWs in multi-band systems. Therefore, the magnetic-field induced phenomena that we report here are not specific to the square lattice TB model, which is chosen here for simplicity.  Materials specific applications of our theory are out of the scope of the present work and are left for future investigations.}

The above described mechanism relies on the dominance of the paramagnetic (Zeeman) effect over the orbital effect, therefore it is most relevant in two-dimensional materials where these conditions can be satisfied by applying the magnetic field in-plane. Moreover, in two-dimensional systems Coulomb interactions are generally enhanced due to the reduced dimenionality \cite{Sharma2018}, thus the tendency for SDW ordering is also enhanced. Our choice of comparable values for the effective potentials $V^{w(m)}$ is in line with this general picture. In fact, several 2D TMDs have been shown to exhibit such competing interactions \cite{Loon2018,Isaacs2016} and it has also been proposed that magnetism and CDW order are closely related \cite{Gueller2016,Law2017,Divilov2020}. {\blue Among them, the strongly-correlated CDW systems 2H-NbSe$_2$ \cite{Divilov2020} and 1T-TaSe$_2$ \cite{Vaskivskyi2016} appear as plausible platforms for the experimental observation of our predicted high-field phases.}
{\blue An experimental platform for the realization of our predicted phases that complies with our here used TB model are the Rare-Earth (RE) tellurides, which are well-known single-Q CDW systems \cite{Kogar2020} and in fact some members exhibit antiferromagnetism, as well  \cite{Chudo2007,Ru2008}. Progress in exfoliating RE-tellurides to the ultrathin limit has been achieved just recently \cite{Xu2020,Lei2020}.}

\section{Conclusions}

In conclusion, we have presented a generalized mean-field theory that allows to study the effect of applied magnetic fields in Pauli limited two-dimensional CDW systems while fully including incommensurability, imperfect nesting and temperature effects and the possibility for a competing/coexisting SDW order. Our numerical solutions showed that the magnetic field -- temperature phase diagram of such systems can contain several different phases depending on the nesting properties of the underlying Fermi surface and the interplay between CDW and SDW ordering and revealed two different mechanisms that could allow the CDW to survive beyond the Pauli paramagnetic limit. For systems with perfect nesting, this can happen through a transition to a FFLO CDW phase that is completely  analogous to the superconducting case. For imperfectly nested systems, near the Pauli critical field the FFLO CDW is unstable against a phase where commensurate CDW and SDW coexist. This phase can appear below the Pauli limit and can therefore be observable for lower fields than $H_P$. Interestingly, in this phase of coexisting CDW+SDW the system is not fully gapped so that a finite magnetization develops similar to the FFLO phase. However, this state is not $q$-modulated and at low temperature it is bounded by two first order metamagnetic transitions. Notably, the CDW state can survive for even higher fields by allowing the coexistence phase to become $q$-modulated. This new high-field phase has characteristics that are a mixture of the commensurate CDW+SDW phase and a FFLO phase, yet it is distinct. Our work reveals that two-dimensional CDW systems can host new exotic high-field phases that go beyond the FFLO paradigm and paves way for their experimental detection.


\bibliographystyle{apsrev4-1}
%

\end{document}